\documentclass[osa,twocolumn,showpacs,floatfix]{revtex4}

\usepackage{here}

\usepackage[dvips]{graphicx}

\newcommand\pictc[5]{\begin{figure}
                       \centerline{
                       \includegraphics[width=#1\columnwidth]{#3}}
                   \protect\caption{\protect\label{fig:#4} #5}
                    \end{figure}            }
\newcommand\pict[4][1.]{\pictc{#1}{!tb}{#2}{#3}{#4}}
\newcommand\rpict[1]{\ref{fig:#1}}

\newcommand\leqt[1]{\protect\label{eq:#1}}
\newcommand\reqtn[1]{\ref{eq:#1}}
\newcommand\reqt[1]{(\reqtn{#1})}

\newcounter{Fig}

\begin{document}
\begin{sloppy}

\title{Discrete gap solitons in modulated waveguide arrays}

\author{Andrey A. Sukhorukov}
\author{Yuri S. Kivshar}

\affiliation{Nonlinear Physics Group,
Research School of Physical Sciences and Engineering,
Australian National University,
Canberra, ACT 0200, Australia}
\homepage{http://www.rsphysse.anu.edu.au/nonlinear}

\begin{abstract}
We suggest a novel concept of diffraction management in waveguide arrays and predict the existence of discrete gap solitons that possess the properties of both conventional discrete and Bragg grating solitons. We demonstrate that both the soliton velocity and propagation direction can be controlled by varying the input light intensity.
\end{abstract}

\ocis{190.4390, %
      190.4420  %
     }

\maketitle

Photonic structures with a periodic modulation of the refractive index (e.g. photonic crystals) can be used to precisely control propagation of optical pulses and beams. Wave localization is possible inside the {\em band gaps} of the linear spectrum, whereas dispersion and diffraction characteristics are strongly modified near the band edges. In particular, recent papers~\cite{Eisenberg:2000-1863:PRL, Morandotti:2001-3296:PRL, Pertsch:2002-93901:PRL} reported on the fabrication of one-dimensional periodic arrays of identical optical waveguides where the effective diffraction coefficient can vanish or even become negative, being controlled by the input conditions and array parameters.

\pict{fig01.eps}{array}{Schematic of a binary array of waveguides with alternating widths $d_1$ and $d_2$ and separation $d_s$.}

In comparison with homogeneous media, efficiency of nonlinear processes can be greatly enhanced in properly designed periodic structures. For waveguide arrays, where waves are primarily localized in weakly coupled waveguides, the effective diffraction can be greatly reduced, further lowering the threshold for the beam self-focusing. It was predicted that diffractional spreading is suppressed for {\em discrete solitons}~\cite{Christodoulides:1988-794:OL, Kivshar:1993-1147:OL}, that are known to possess many remarkable properties~\cite{Lederer:2001-269:SpatialOptical}. For example, unlike their continuous counterparts, discrete solitons can propagate across an array at low powers, while at high powers they become trapped by array discreteness~\cite{Claude:1993-14228:PRB, Aceves:1996-1172:PRE}, and this behavior is readily observed in experiment~\cite{Morandotti:1999-2726:PRL}. It was also demonstrated that discrete solitons can be efficiently routed through two-dimensional networks of coupled waveguides~\cite{Christodoulides:2001-233901:PRL}.

In a number of recent studies~\cite{Ablowitz:2001-254102:PRL, Peschel:2002-544:JOSB}, it was suggested that the properties of discrete solitons can be modified by a periodic modulation of waveguides along the propagation direction. In this Letter, we suggest another, even more fundamental concept of the array engineering and consider {\em a binary waveguide array} composed of alternating ``thick'' and ``thin'' waveguides, as illustrated in Fig.~\rpict{array}. In such a structure, the effective refractive index experiences additional transverse modulation and, therefore, a ``Rowland ghost gap'' may appear in the linear spectrum~\cite{Russell:1986-596:PRL}. Formation of solitons in such gaps was earlier studied in the context of superstructure fiber Bragg gratings~\cite{Broderick:1995-5788:PRE}, where the analysis is based on the coupled-mode equations. In this Letter, we demonstrate, for the first time to our knowledge, the existence of {\em discrete gap solitons} that display the properties of both conventional discrete and Bragg grating solitons and resemble nonlinear localized gap modes in diatomic lattices~\cite{Kivshar:1992-7972:PRA}.

Propagation of waves in an array of weakly coupled single-mode
optical waveguides can be approximately
described~\cite{Christodoulides:1988-794:OL, Kivshar:1993-1147:OL} by the discrete nonlinear Schr\"odinger equation (DNLS) for the normalized
amplitude of the electric field $E_n$ localized at the waveguide
with the index $n$,
\begin{equation} \leqt{DNLS}
   i \frac{d E_n}{dz}
   + \lambda_n E_n
   + (E_{n-1} + E_{n+1})
   + \gamma_n |E_n|^2 E_n
   = 0,
\end{equation}
where $\lambda_n$ characterizes the linear propagation constant of
the mode guided by the $n-$th waveguide (which depends on its
width), and $\gamma_n$ is the effective nonlinear coefficient. For
the analysis of the structure shown in Fig.~\rpict{array}, it is
convenient to introduce the notations: $a_n = E_{2 n}$, $b_n = E_{2
n + 1}$, $\lambda_{2 n + 1} \equiv - \rho$, and $\lambda_{2 n}
\equiv 0$, assuming the appropriate normalization of
Eq.~\reqt{DNLS}. In order to simplify the analysis, we neglect
absorption and also consider the identical nonlinear coefficients,
$\gamma_n \equiv \gamma$. However, we have verified that the main
conclusions of our study remain valid if $\gamma_n$ are weakly
modulated. With no loss of generality, we have $\rho>0$, so that
$a_n$ and $b_n$ are the field amplitudes at the thick and thin
waveguides, respectively. Dependence of the normalized detuning on
the waveguide widths $d_{1,2}$ and the separation
$d_s$ (see Fig.~\rpict{array}) can be found numerically.

\pict{fig02.eps}{dispers}{Characteristic dependencies of the
Bloch wave number ($K_b$), group velocity ($v_g$), and the
effective diffraction coefficient ($D$) on the propagation
constant $\beta$. Gray shadings mark the transmission bands.
Normalized detuning between the thick and thin waveguides is $\rho
= 1.5$. }

Our structure can be compared to the waveguide arrays with defects
\cite{Peschel:1999-1348:APL}, where localization at a thin waveguide ($d_2=2.5\,{\rm \mu m}$) embedded in an array of thicker waveguides ($d_1=4\,{\rm \mu m}$) with $d_s=5\,{\rm \mu m}$ was recently observed in experiments with AlGaAs arrays. Under such
conditions, the normalized detuning between the modes of the thin
and thick waveguides is about $\rho \simeq 1.5$, the value we use
in numerical simulations presented below. Defects with other parameters can also be fabricated~\cite{Morandotti:2002-239:ProcCLEO}.

First we analyze the properties of {\em linear Bloch waves} in
such a periodic binary structure. The Bloch waves are
characterized by the wave number $K_b$,
\begin{equation} \leqt{Bloch}
   a_n = A e^{i \beta z + i K_b n}, \quad b_n = B e^{i \beta z + i K_b n},
\end{equation}
where $\beta$ is the mode propagation constant. We substitute
Eq.~\reqt{Bloch} into the linearized equation \reqt{DNLS} (with
$\gamma=0$), and obtain the {\em linear dispersion relation} that
couples the propagation constant and Bloch wave number, and
yields a relation between the amplitudes at the thin and thick
waveguides,
\begin{equation} \leqt{dispers}
   K_b = \cos^{-1}( \eta / 2) , \quad
   A \beta = B e^{-i K_b/2} \sqrt{2 - \eta},
\end{equation}
where $\eta = 2 - \beta (\beta+\rho)$. It follows that the transmission bands, corresponding to real $K_b$, appear when $\beta_- \le \beta \le -\rho$ or $0 \le \beta \le \beta_+$, where $\beta_\pm = -(\rho/2) \pm \sqrt{(\rho/2)^2+4}$. On the other hand, an additional gap, also called the Rowland ghost gap~\cite{Russell:1986-596:PRL}, appears for $-\rho < \beta < 0$, and it increases for a larger difference between the widths of the neighboring waveguides. A characteristic dispersion relation and the corresponding band-gap structure are presented in Fig.~\rpict{dispers}(a).

Using the dispersion relation~\reqt{dispers}, we calculate the {\em group velocity}, $v_g = - 2 (\partial \beta / \partial K_b)$, and the effective {\em diffraction coefficient}, $D = - 4 (\partial^2 \beta / \partial K_b^2)$, of the binary array. Characteristic dependencies of these parameters on the propagation constant $\beta$ are shown in Figs.~\rpict{dispers}(b,c). In both transmission bands, there exist regions with effective normal and anomalous diffraction separated by a zero-diffraction point.

When the medium possesses a self-focusing nonlinearity, bright discrete solitons can form provided the effective diffraction coefficient is positive. We consider excitation of such solitons by an input Gaussian beam,
\begin{equation} \leqt{Gauss}
   E_n(z=0) = C \exp\left[-(n-n_0)^2 / \nu^2 + i \kappa (n-n_0) \right] ,
\end{equation}
where $n_0$ is the position of the beam center, $\nu$ is the beam width, and $\kappa$ characterizes the inclination angle. Such an input beam can be presented as a superposition of two modulated linear eigenmodes, whose eigenvalues $\beta_{1,2}$ correspond to the Bloch wave number $K_b = \cos^{-1}[\cos(2 \kappa)]$,
\[
   \begin{array}{l} {\displaystyle
      E_{2 n}(0) = (A_1 + A_2)
                  \exp\left[-(2 n-n_0)^2 / \nu^2 + i K_b n \right] ,
   } \\*[9pt] {\displaystyle
      \,
      E_{2 n+1}(0) = (B_1 + B_2)
                  \exp\left[-(2 n+1-n_0)^2 / \nu^2 + i K_b n \right] ,
   } \end{array}
\]
where the amplitudes $A_j$ and $B_j$ ($j=1,2$) satisfy
Eq.~\reqt{dispers} at $\beta=\beta_j$ (we choose $\beta_1 > \beta_2$, with no loss of generality). For a wide input beam ($\nu \ge
4$), a ratio of the powers of two excited Bloch modes can be
written in a simple form,
\begin{equation}
   p = \frac{P_2}{P_1} \simeq \left( \frac{1-\delta}{1+\delta} \right)^2 ,
\end{equation}
where $\delta = \left[\sqrt{1 + \rho / \beta_1} \cos\kappa\right] / \cos(K_b/2)$. The two modes always have opposite group velocities and diffraction coefficients (see Fig.~\rpict{dispers}).

\pict{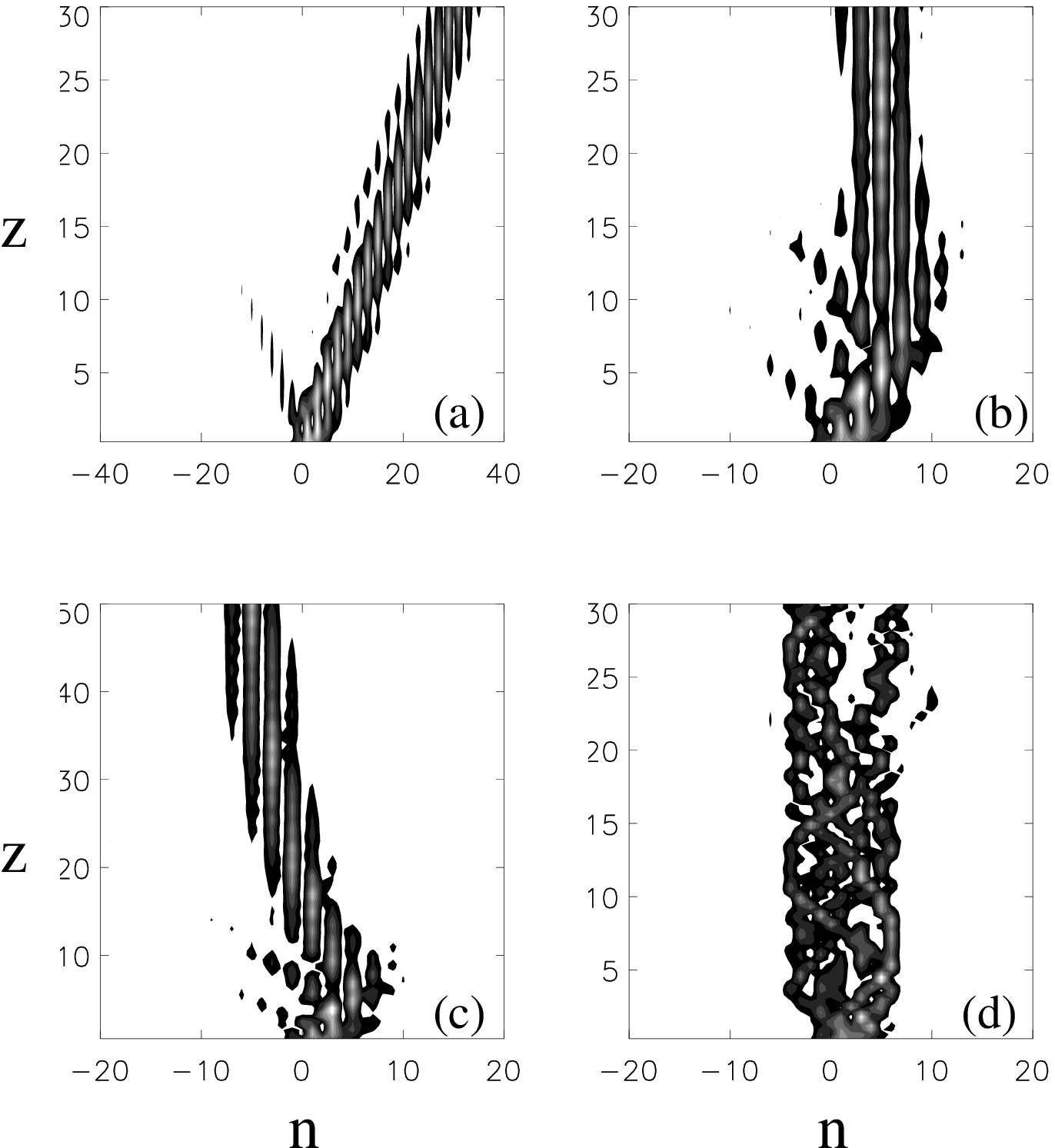}{bpm}{Excitation of discrete gap solitons in a
self-focusing medium ($\gamma=+1$) by input beams with $\kappa =
0.6 \pi$, $n_0=1$, $\nu = 4$, and normalized peak intensities
(a)~$|C|^2=0.3$, (b)~$0.75$, (c)~$0.9$, (d)~$5$. The parameters correspond to Fig.~\rpict{dispers}. }

It is useful to compare the results with a homogeneous array, when $\rho=0$. In this case, within the validity of the DNLS model, there exists only one parameter domain where diffraction is positive, and discrete solitons can exist, for self-focusing media~\cite{Eisenberg:2000-1863:PRL, Morandotti:2001-3296:PRL, Pertsch:2002-93901:PRL}. On the other hand, in a modulated array diffraction is modified close to the Rowland ghost gap, and there appear two regions with positive diffraction, where solitons can exist. Thus, the binary array structure allows the band gap engineering in a broad parameter region.

Let as first discuss the excitation of discrete solitons by a beam
incident at a normal angle, so that $\kappa = K_b = 0$,
and $\beta_{1,2} = \beta_{+,-}$. In this case, $p < 1$, i.e. the
first ($j=1$) mode is always dominant (for example, $p \simeq
0.03$ for $\rho = 1.5$). As is demonstrated in Fig.~\rpict{dispers}, at
$\beta=\beta_+$ the effective diffraction coefficient is positive
and, therefore, discrete solitons can be formed in a self-focusing
medium. Properties of such solitons are similar to those
in homogeneous arrays.

On the other hand, {\em discrete gap solitons} may appear due to localization withing the linear gap (at $-\rho<\beta<0$), which appears due to a peridic modulation of waveguide widths as discussed above.
In order to excite stationary gap solitons in a self-focusing
medium, one might attempt to use an input beam with $\kappa =
\pi/2$, so that $K_b = \pi$, $\beta_1 = 0$, $\beta_2 = -\rho$. In
this case, two Bloch modes have equal powers ($p=1$), however
one of them experiences self-focusing and another one~---
self-defocusing, and in this situation an efficient generation of
gap solitons is not possible.

The optimum conditions for generating discrete gap solitons can be
realized when $\pi/2 < \kappa < \pi$, and $D(\beta_2) > 0$.
Indeed, under such conditions $p>1$, i.e. the second mode, which
experiences self-focusing, is dominant. Moreover, the Bloch wave
envelopes have opposite group velocities, so that they move apart
in the opposite directions. In Fig.~\rpict{bpm}(a), we show that
two beams are indeed generated at the input. The beam which moves
to the right is localized at odd (i.e. thin) waveguides, and it
transforms into a gap soliton. On the contrary, the other beam
moves to the left, and it experiences self-defocusing and broadens.

Stationary gap solitons can be generated at higher input
intensities. Under such conditions, two Bloch modes are
initially trapped together, resulting in a periodic beating that
can strongly affect the soliton formation process. If the emerging
soliton has a small velocity, it is not able to overcome the
Peierls-Nabarro potential of the periodic structure, and it becomes
trapped~\cite{Aceves:1996-1172:PRE}. In this case, the gap soliton
can remain in the center, as is shown in Fig.~\rpict{bpm}(b).
However, if the soliton acquires a higher velocity, it can still
move across the array, as is shown in Fig.~\rpict{bpm}(c). Somewhat
surprisingly, in the latter case the soliton moves to the left,
i.e. in the direction opposite to the propagation direction of the
input beam. If the intensity is increased even further, then the
gap solitons become oscillatory unstable, similar to the solitons
in fiber Bragg gratings~\cite{Barashenkov:1998-5117:PRL} [see Fig.~\rpict{bpm}(d)].

In conclusion, we have studied the diffraction properties and
nonlinear wave propagation in binary waveguide arrays with
alternating waveguide widths. We have predicted the existence of
discrete gap solitons and demonstrated their intriguing dynamics
controlled by varying the input intensity.

The authors thank J.~S. Aitchison, R. Morandotti, and Y.~Silberberg for useful discussions.
This work was supported by the Australian Research Council.

\end{sloppy}

\begin{thebibliography}{10}

\bibitem{Eisenberg:2000-1863:PRL}
H.~S. Eisenberg, Y. Silberberg, R. Morandotti, and J.~S. Aitchison,
  ``Diffraction management,'' Phys. Rev. Lett. {\bf 85,} 1863--1866 (2000).

\bibitem{Morandotti:2001-3296:PRL}
R. Morandotti, H.~S. Eisenberg, Y. Silberberg, M. Sorel, and J.~S. Aitchison,
  ``Self-focusing and defocusing in waveguide arrays,'' Phys. Rev. Lett. {\bf
  86,} 3296--3299 (2001).

\bibitem{Pertsch:2002-93901:PRL}
T. Pertsch, T. Zentgraf, U. Peschel, A. Brauer, and F. Lederer, ``Anomalous
  refraction and diffraction in discrete optical systems,'' Phys. Rev. Lett.
  {\bf 88,} 093901--4 (2002).

\bibitem{Christodoulides:1988-794:OL}
D.~N. Christodoulides and R.~I. Joseph, ``Discrete self-focusing in nonlinear
  arrays of coupled wave-guides,'' Opt. Lett. {\bf 13,} 794--796 (1988).

\bibitem{Kivshar:1993-1147:OL}
Yu.~S. Kivshar, ``Self-localization in arrays of defocusing wave-guides,'' Opt.
  Lett. {\bf 18,} 1147--1149 (1993).

\bibitem{Lederer:2001-269:SpatialOptical}
F. Lederer, S. Darmanyan, and A. Kobyakov, ``Discrete solitons,''  in {\em
  Spatial Optical Solitons}, Vol.~82 of {\em Springer series in optical
  sciences}, S. Trillo and W.~E. Torruellas, eds., (Springer-Verlag, New York,
  2001), \ pp.\ 269--292.

\bibitem{Claude:1993-14228:PRB}
C. Claude, Yu.~S. Kivshar, O. Kluth, and K.~H. Spatschek, ``Moving localized
  modes in nonlinear lattices,'' Phys. Rev. B {\bf 47,} 14228--14232 (1993).

\bibitem{Aceves:1996-1172:PRE}
A.~B. Aceves, C. {De Angelis}, T. Peschel, R. Muschall, F. Lederer, S. Trillo,
  and S. Wabnitz, ``Discrete self-trapping, soliton interactions, and beam
  steering in nonlinear waveguide arrays,'' Phys. Rev. E {\bf 53,} 1172--1189
  (1996).

\bibitem{Morandotti:1999-2726:PRL}
R. Morandotti, U. Peschel, J.~S. Aitchison, H.~S. Eisenberg, and Y. Silberberg,
  ``Dynamics of discrete solitons in optical waveguide arrays,'' Phys. Rev.
  Lett. {\bf 83,} 2726--2729 (1999).

\bibitem{Christodoulides:2001-233901:PRL}
D.~N. Christodoulides and E.~D. Eugenieva, ``Blocking and routing discrete
  solitons in two-dimensional networks of nonlinear waveguide arrays,'' Phys.
  Rev. Lett. {\bf 87,} 233901--4 (2001).

\bibitem{Ablowitz:2001-254102:PRL}
M.~J. Ablowitz and Z.~H. Musslimani, ``Discrete diffraction managed spatial
  solitons,'' Phys. Rev. Lett. {\bf 87,} 254102--4 (2001).

\bibitem{Peschel:2002-544:JOSB}
U. Peschel and F. Lederer, ``Oscillation and decay of discrete solitons in
  modulated waveguide arrays,'' J. Opt. Soc. Am. B {\bf 19,} 544--549 (2002).

\bibitem{Russell:1986-596:PRL}
P.~St.~J. Russell, ``Bragg resonance of light in optical superlattices,'' Phys.
  Rev. Lett. {\bf 56,} 596--599 (1986).

\bibitem{Broderick:1995-5788:PRE}
N.~G.~R. Broderick, C.~M. {de Sterke}, and B.~J. Eggleton, ``Soliton solutions
  in Rowland ghost gaps,'' Phys. Rev. E {\bf 52,} R5788--R5791 (1995).

\bibitem{Kivshar:1992-7972:PRA}
Yu.~S. Kivshar and N. Flytzanis, ``Gap solitons in diatomic lattices,'' Phys.
  Rev. A {\bf 46,} 7972--7978 (1992).

\bibitem{Peschel:1999-1348:APL}
U. Peschel, R. Morandotti, J.~S. Aitchison, H.~S. Eisenberg, and Y. Silberberg,
  ``Nonlinearly induced escape from a defect state in waveguide arrays,'' Appl.
  Phys. Lett. {\bf 75,} 1348--1350 (1999).

\bibitem{Morandotti:2002-239:ProcCLEO}
R. Morandotti, H.~S. Eisenberg, D. Mandelik, Y. Silberberg, D. Modotto, M.
  Sorel, and J.~S. Aitchison, ``Interactions of Discrete Solitons with Defects
  and Interfaces,'' In {\em Quantum Electronics and Laser Science Conference,
  OSA Technical Digest}, {Postconference} ed.\ OSA Trends in Optics and
  Photonics (TOPS)\ {\bf 74,} 239  (Optical Society of America, Washington DC,
  2002).

\bibitem{Barashenkov:1998-5117:PRL}
I.~V. Barashenkov, D.~E. Pelinovsky, and E.~V. Zemlyanaya, ``Vibrations and
  oscillatory instabilities of gap solitons,'' Phys. Rev. Lett. {\bf 80,}
  5117--5120 (1998).

\end{thebibliography}
\end{document}